\documentclass[a4paper]{article}
\usepackage{amssymb,amsthm,amscd, amsbsy, array}
 \usepackage{amsmath,bm,cite,color}
 \usepackage{graphicx,float,bm}
\def\eq#1{\begin{equation}#1\end{equation}}

\def\eqs#1{\begin{eqnarray}#1\end{eqnarray}}
\def\bea{\begin{eqnarray}}
\def\eea{\end{eqnarray}}
\newcommand{\lf}{\left (}
\newcommand{\rg}{\right )}

\author{D. Levi\footnote{e-mail:
Decio.Levi@roma.infn..it},$\quad$  L. Martina\footnote{e-mail:
Luigi.Martina@le.infn.it}, $\quad$ P. Winternitz\footnote{e-mail:
wintern@crm.umontreal.ca}
\\
 $*$   Dipartimento di Matematica e Fisica, Universit\'a di  Roma Tre \\and  INFN, Sezione di Roma Tre \\ Via della Vasca Navale 84, 00146 Roma, Italy\\
\dag Dipartimento di Matematica e Fisica, Universit\'a del Salento\\  and INFN, Sezione di  Lecce  \\
Via per Arnesano, C.P. 193 I-73100 Lecce, Italy\\
\ddag D\'epartement de Math\'ematiques et de Statistique and \\ Centre de Recherches Math\'ematiques,\\
Universit\'e de Montr\'eal, C.P. 6128, \\succ. Centre-ville, Montr\'eal (QC) H3C 3J7,
Canada
 }
 \title{Conformally invariant elliptic Liouville equation and its symmetry preserving discretization.}

\begin{document}
\maketitle

\begin{abstract}
 The symmetry algebra of the real elliptic Liouville equation is an
infinite-dimensional loop algebra with the simple Lie algebra $o(3,1)$ as
its maximal finite-dimensional subalgebra. The entire algebra generates
the conformal group of the Euclidean plane $E_2$. This infinite-dimensional
algebra distinguishes the elliptic Liouville equation from the hyperbolic
one with its symmetry algebra that is the direct sum of two Virasoro
algebras. Following a discretisation procedure developed earlier, we
present a difference scheme that is invariant under the group $O(3,1)$ and
has the elliptic Liouville equation in polar coordinates as its continuous
limit. The lattice is a solution of an equation invariant under $O(3,1)$ and
is itself invariant under a  subgroup  of $O(3,1)$, namely the $O(2)$ rotations of the
Euclidean plane.

  \end{abstract}
\section{Introduction}
The Liouville equation is 
\bea \label{1.1}
z_{\alpha\beta}=e^z,
\eea
where $z,\,\alpha$, and $\beta$ are complex variables. Its two inequivalent  real forms are obtained when  $z,\,\alpha$, and $\beta$ are real (the hyperbolic Liouville equation), or when $\beta$ is the complex conjugate of $\alpha$  (the elliptic Liouville equation). They  are both among the most important and ubiquitous equations in physics.   In particular, the  Liouville equation  defines the conformal factor of the metric on a surface of constant Gaussian curvature \cite{DNF}, so  this equation and its quantized version, are of primary importance in gravity theory \cite{Jackiw,k,bpz,p87,Polyakov,Zamolodchikov,Dorn, Teschner,Nakayama} and in several gauge field models \cite{JaffeTaubes,Tarantello}.  Its general  solution has been known for a long time and was discussed in detail in \cite{Crowdy}.

    In two recent articles \cite{4,5} we  dealt with the real hyperbolic Liouville equation as part of a program of discretizing  both ordinary and partial differential equations   (ODEs and PDEs),  while preserving their Lie point symmetries \cite{doro, lw}.   In particular we have shown that it is  not possible  to preserve the entire infinite-dimensional symmetry group as point symmetries in the discrete case.    The Lie symmetry algebra  in the  hyperbolic Liouville equation case  is the direct sum of two Virasoro algebras  $ vir_x  \bigoplus vir_y $.   Our discretization preserves the maximal finite-dimensional subalgebra, namely  $ sl_x \lf 2 , \mathbb{R} \rg \bigoplus sl_y \lf 2 , \mathbb{R} \rg $.  An alternative approach  \cite{6,7}   to ours makes it possible to preserve  the entire  direct product $ VIR\lf x \rg \otimes VIR\lf y \rg$  group,  however as generalized symmetries
rather than point ones.
 A similar pattern emerges in the symmetry preserving discretization   of the elliptic Liouville equation. As we will see, it is possible to preserve only the maximal finite-dimensional subalgebra as point symmetries, not  the entire algebra. New features appear in the elliptic case. The symmetry algebra in the continuous case is simple, rather than semisimple. In the discretization the lattice is invariant under rotations, rather than translations as in the hyperbolic case.

Let us  briefly review the role that  infinite-dimensional Lie point symmetry algebras play for various types of differential equations and why we would like to preserve them, as much as possible in discretizing procedures. The existence of an infinite-dimensional symmetry group is very often an indication of some sort of integrability. By this we mean the possibility of obtaining all solutions, or large classes of solutions, by essentially linear techniques.

For ODEs  the only equations that allow an infinite-dimensional algebra are first order ones. Finding  any one-dimensional subalgebra of this infinite-dimensional Lie algebra is equivalent to finding an integrating factor. It was shown in \cite{rw} that a discretization of a first order ODE preserving a one-dimensional symmetry algebra is exact. The obtained invariant two-point difference scheme has exactly the same solutions as the original ODE.

For nonlinear PDEs the existence of an infinite-dimensional Abelian Lie point symmetry group of a specific type is an indication of linearizability by an invertible transformation of variables \cite{bk,gls}. The Abelian Lie algebra is just a reflection of the linear superposition principle for the linear equations.

Finally, all nonlinear PDEs involving three independent variables that are solvable by the Inverse Scattering Method have infinite-dimensional symmetry algebras with a Kac--Moody--Virasoro structure \cite{dklw1,dklw2,cw,pw,mw,ow,dlw}.

For classical studies of PDEs with infinite-dimensional Lie point symmetry algebras and their classification see \cite{1,2,ama}. An interesting comment made by Lie in \cite{1} is the statement that the integration of any second order PDE in two independent variable which has an infinite-dimensional symmetry group is reduced to the integration of three ODEs. This is related to Darboux integrability \cite{gls}.

First order linear or linearizable delay ordinary differential equations \newline  (DODEs) typically do have infinite-dimensional symmetry algebras. For their discretization see \cite{dkmw}.

The discretization of  PDEs and DODEs preserving their Lie point symmetry groups started in \cite{4,5,25,41,42,43,dkmw,ltw} and the present article is an integral part of this research.  As stated in previous articles there are several aspects to the program. The most fundamental is that the world may  be inherently discrete (for instance because of the Planck length). Then differential equations would be approximations of difference ones. In quantum field theory discretizations are part of renormalization. Keeping conformal, Lorentz or Galilei symmetries is an important requirement \cite{lpw}. Some physical phenomena are discrete, even if space-time is continuous (crystals, molecular or atomic chains, spin chains, $\cdots$). Finally, discretizations are an essential part of  numerical algorithms for solving PDEs, DODEs or ODEs.

This last aspect places our program in the field of geometric integration \cite{11, 14, 26, 27}. That is an attempt to preserve some important features of the physical problem in the discretization. This feature may be its Hamiltonian or symplectic structure, some or all integrals of motion or  known asymptotic behaviour, etc. . In our case we wish to preserve, whenever possible, the entire Lie point symmetry group.  When that is not possible we preserve at least some maximal subgroup of the symmetry group. The motivation for this is that the symmetry group  of a PDE, a DODE,  an ODE  or a system of such equations encodes a large amount of information about the solution set of the equations.  Moreover, symmetries are of fundamental importance in physics and it is a pity to loose them when one starts to do numerics.  Most symmetries are lost as soon as any specific fixed lattice is postulated (e.g. a Cartesian lattice cannot be invariant under rotations, a frequent symmetry of a physical system).                                                                                                                                                                                                                                                                                                                                                                                                                                                                   This may  be avoided by including the lattice variables as new dynamical  fields, determined by the continuous symmetry algebra, as we will show below.   The proposed approach is to replace a PDE 
\bea \label{a2}
\mathcal E(x, y, u, u_x, u_y, u_{xx}, u_{xy}, u_{yy}, \cdots)=0,
\eea
by a set of relations between points
\bea \label{a3}
&&\mathcal E_a \Big (x_{m+i,n+j}, y_{m+i,n+j}, u(x_{m+i,n+j},y_{m+i,n+j}) \Big)=0, \\ \nonumber
&& 1 \le a \le N, \quad 0 \le m \le N_1, \quad 0 \le n \le N_2,\quad 0 \le i \le k_1, \quad 0 \le j \le k_2
\eea
where $m$ and $n$ specify a stencil  and $i$ and $j$ the position on the stencil. The numbers $N$  (in our case $N=3$) ,  $N_1$, $N_2$,  $k_1$ and $k_2$ depend on the number of partial derivatives present in \eqref{a2} .  In the continuous limit the constraint is that some combination of the relations $\mathcal E_a=0$ \eqref{a3} should go into the PDE \eqref{a2}, the other relations into identities (like $0=0$).

We call the difference scheme \eqref{a3} invariant when it is constructed entirely out of invariants of the Lie point symmetry  group $ G$ of the PDE \eqref{a2}, or some specific subgroup $ G_0 \subset  G$. Thus both the (approximate) solution of the PDE \eqref{a2} and the lattice emerge as solutions of \eqref{a3}. 

To avoid possible confusion we immediately state that invariance of equations under some group $ G$ does not imply invariance of solutions. Solutions also satisfy boundary or initial conditions (or some combination of both) that break the symmetry. What is invariant is the set of all solutions.  To obtain numerical solutions the lattice must be specified completely. In particular, we  choose the lattice to be a solution of the  invariant scheme \eqref{a3}. After the choice is made (compatible with \eqref{a2}) the lattice is no longer invariant. Similarly, once we impose boundary conditions and start integrating, we obtain a non-invariant solution  (transformed into another solution of \eqref{a3} by the its Lie point symmetry group).

The  method of preserving Lie point symmetries sketched above has been already  proved to be useful, for instance  when doing numerics.   For ODEs invariant methods provide precise solutions close to singularities and continued beyond singularities, where conventional Runge-Kutta methods break down \cite{a1,a2,a3,a4,a5,a6,a7,a8}. For PDEs we refer to our previous articles on the hyperbolic  Liouville equation \cite{4,5}. We compared, under identical conditions, four different numerical methods. A standard one applicable to virtually any PDE, one preserving the entire infinite-dimensional symmetry group as generalized symmetries \cite{6,7}, our method preserving the maximal finite-dimensional  subgroup as point symmetries  \cite{4,5}   and one preserving linearizability \cite{3}. In the comparison the last two performed  about two orders of magnitude better then the first two. For other numerical applications of symmetry preserving schemes to PDEs  and comparison with standard methods see e.g. \cite{bv}.

The structure of this article is as follows. In Section 2 we write the complex algebraic Liouville equation and its Lie point symmetry algebra, the direct sum of two complex Virasoro algebras.  We then restrict to the real elliptic Liouville equation and transform to polar coordinates. The obtained real symmetry algebra is an infinite-dimensional simple Lie algebra and its maximal finite-dimensional subalgebra is identified as $ O(3,1)$. In Section 3 we introduce a 9-point stencil labeled by a reference point $(m,n)$ and involving the points $(m+i,n+j)\;$, $0 \le i \le 2\;$, $0 \le j \le 2$. On this lattice we construct a total of 24 complex invariants, each depending on 4 points only. Using these  $ SL\lf 2,\mathbb C \rg \otimes SL\lf 2,\mathbb 
C \rg$ invariants we discretize the complex Liouville equation. We then restrict appropriately to the real elliptic equation, to real variables and real $O(3,1)$ invariants.
We express these real  invariants in polar coordinates and construct a polar type lattice, on which we write a discrete elliptic Liouville equation which has the real elliptic Liouville equation in polar coordinates  \eqref{2.9}  as its continuous limit.  Section 4 is devoted to a short conclusion and to a future outlook.

\section{The complex Liouville equation, its Lie point symmetry algebra and their real forms}
The {\bf complex Liouville equation}  (\ref{1.1}) has  the algebraic form
\bea \label{2.2}
u u_{\alpha \beta} - u_{\alpha} u_{\beta} = u^3, \qquad u=e^z,
\eea
where $z$ and $u$ are (complex) functions of two complex variables $\alpha$ and $\beta$.

Its Lie point symmetry algebra is known (or at least it can be easily deduced from classical studies of partial differential equations with infinite-dimensional symmetry groups \cite{ama,1,2}) and from more recent articles on the real Liouville equation \cite{9,10}. It is the direct sum of two Lie algebras, each isomorphic to the algebra of diffeomorphisms of the complex line (a.k.a. the complex Virasoro algebra). For the algebraic Liouville equation (\ref{2.2}) the symmetry algebra is realized by the holomorphic vector fields
\bea \label{2.3}
A(f(\alpha))=f(\alpha) \partial_{\alpha} - f'(\alpha)\,u\,\partial_u, \\
B(g(\beta))=g(\beta) \partial_{\beta} - g'(\beta)\,u\,\partial_u.
\eea
Here $f(\alpha)$ and $g(\beta)$ are arbitrary holomorphic functions of $\alpha$ and $\beta$ respectively (but not of the complex conjugates $\bar \alpha$ and $\bar \beta$).

The most important mathematical properties of the Liouville equation (\ref{1.1}) (and  (\ref{2.2})) are
\begin{enumerate}
\item The Lie point symmetry group is infinite-dimensional.
\item Equation (\ref{1.1}) is linearized by a transformation of the dependent variable
\bea \label{2.4}
z= \ln (2 \frac{\phi_{\alpha} \phi_{\beta}}{\phi^2}), \qquad \phi_{\alpha \beta} =0.
\eea
\end{enumerate}

Thus, every solution of the complex Laplace equation (\ref{2.4}) ($\phi=\phi_1(\alpha) + \phi_2(\beta)$ with $\phi_1$ and $\phi_2$ arbitrary functions of their argument) provides a solution of the Liouville equation (\ref{1.1}).

The  real hyperbolic Liouville equation and its symmetry algebra are obtained in exatly the form (\ref{1.1}),$\dots$, (\ref{2.4}) by taking $\alpha=x$, $\beta=y$ and considering $\{x, y, z\}$ to be real.
 
The { real elliptic Liouville equation} is obtained by putting
\bea \nonumber
\alpha=\frac 1 2 (x + i y), \qquad \beta=\bar \alpha= \frac 1 2 (x - i y), \qquad \{x,y,z\} \in \mathbb R^3
\eea
to  get from (\ref{1.1})
\bea \label{2.5}
z_{xx} + z_{yy} = e^z,
\eea
and from (\ref{2.2})
\bea \label{2.6}
u (u_{xx} + u_{yy})- (u_x^2+u_y^2) = u^3, \qquad u=e^z.
\eea
Its Lie point symmetry algebra can be written as 
\bea \label{2.7}
\hat X = \xi(x,y) \partial_x + \eta(x,y) \partial_y - (\xi_x + \eta_y)u \partial_u,
\eea
where the functions $\xi(x,y)$ and $\eta(x,y)$ satisfy  \cite{9,10}
\bea \label{2.8}
\xi_{xx}+ \xi_{yy} = 0, \quad \eta_{xx}+\eta_{yy}=0,
\quad \eta_x=-\xi_y, \quad \eta_y =\xi_x.
\eea 
Thus, $\xi$ and $\eta$ are any two harmonically conjugate real smooth functions.

To make the Lie algebra  explicit  we transform to polar coordinates, rewriting (\ref{2.6})  as
\bea \label{2.9}
u(u_{rr}+\frac 1 r u_r + \frac 1 {r^2} u_{\theta \theta}) - (u_r^2 + \frac 1 {r^2} u_{\theta}^2) = u^3, \quad x=r \cos \theta, \; y = r \sin \theta.
\eea
The equations  (\ref{2.8}) transform into
\bea \label{2.10}
&\xi_{rr} + \frac 1 r \xi_r + \frac 1 {r^2} \xi_{\theta \theta} = 0, \qquad  &\eta_r = - \frac 1 r \xi_{\theta},\\ \nonumber
&\eta_{rr} + \frac 1 r \eta_r + \frac 1 {r^2} \eta_{\theta \theta} = 0, \qquad  &\eta_{\theta} =  r \xi_{r}.
\eea
By separating the variables we obtain
\bea \label{2.11}
\xi(r,\theta) = \sum_{m=0}^{\infty} r^m (a_m \cos m \theta + b_m \sin m \theta), \\ \nonumber
\eta(r, \theta) = \sum_{m=0}^{\infty} r^m (a_m \sin m \theta - b_m \cos m \theta).
\eea
A basis for the Lie point symmetry algebra (\ref{2.7}) can be chosen to be:
\bea \label{2.12}
 A_m &=&  r^m \cos (m-1) \theta\, \partial_r + r^{m-1} \sin (m-1)\theta \, \partial_{\theta} -\\ \nonumber &-&2 m r^{m-1} u\, \cos (m-1)\theta \,\partial_u, \\ \nonumber
 B_m &=& r^m \sin (m-1) \theta\, \partial_r - r^{m-1} \cos (m-1)\theta \,  \partial_{\theta} -\\ \nonumber &-&2 m r^{m-1}\, u \sin (m-1)\theta  \, \partial_u, 
\eea
where $ m \in \mathbb Z^{\ge 0}$.
It should be emphasized that, contrary to the case of the hyperbolic Liouville equation, this infinite-dimensional Lie algebra is not a direct sum of two algebras. In particular, if we restrict to $m=0,1$ and $2$ we obtain the maximal finite-dimensional subalgebra
\bea 
A_0&=&P_1=\cos \theta \partial_r -\frac 1 r \sin \theta \partial_{\theta}, \nonumber \\
 B_0&=&P_2=\sin \theta \partial_r + \frac 1 r \cos \theta \partial_{\theta},  \nonumber \\
B_1&=&L_3=\partial_{\theta}, \quad A_1=D= r \partial_r -2 u \partial_u, \label{2.13} \\ \nonumber
A_2&=&C_1=r(r\,\cos \theta  \partial_r + \sin \theta \partial_{\theta} -4 u\, \cos \theta  \partial_u), \\ \nonumber
B_2&=&C_2=r(r\,\sin \theta  \partial_r - \cos \theta \partial_{\theta} -4 u\, \sin \theta  \partial_u). 
\eea
This is the (simple) algebra $ o(3,1)$, realized as the Lie algebra of the conformal group of the Euclidean plane (a real form of the semisimple complex Lie algebra $ o(4, \mathbb C) \equiv  o(3, \mathbb C) \bigoplus  o(3, \mathbb C)$).  The auxiliary notation introduced in (\ref{2.13}) has a mnemonic value, meaning $P_i$ for translations, $L_3$ rotation, $D$ for dilation and $C_i$ for special conformal generators.

More generally we obtain a simple  real form of the direct sum of two complex Virasoro algebras with commutation relations
\bea \label{2.14}
[A_m, A_{\ell}]&=&(\ell - m) A_{m + \ell -1}, \quad [B_m, B_{\ell}]=(m - \ell) A_{m + \ell -1},\\ \nonumber
[A_m, B_{\ell}] &=& (\ell - m) B_{m+\ell -1}.
\eea   
\section{Invariants in discrete space and  the invariant discrete elliptic Liouville equation}
 The discretization of the hyperbolic Liouville equation with preservation of either its symmetries, or its linearizability was the topic of earlier articles \cite{3,4,5,6,7}. In particular, in the articles \cite{4,5} we constructed a discrete hyperbolic Liouville equation invariant under the maximal finite subgroup $ SL_x \lf 2 , \mathbb{R} \rg \otimes SL_y \lf 2 , \mathbb{R} \rg $. We then showed that it is not possible to extend the invariance to the full group $ VIR\lf x \rg \otimes VIR\lf y \rg$ while restricting to point transformations, { if we want to obtain the hyperbolic Liouville equation in the continuous limit. On the other hand it is easy to discretize the linear wave equation in a completely invariant manner (with an infinite-dimensional symmetry group \cite{ltw})}. In \cite{4} we considered a 4-point stencil, in \cite{5} also a 9-point one and arrived at the same conclusions.  
 Namely, it is not possible to construct a $ VIR\lf x \rg \otimes VIR\lf y \rg$ invariant  discrete invariant
scheme for the hyperbolic Liouville equation out of the 24 $ SL_x \lf 2 , \mathbb{R} \rg \otimes SL_y \lf 2 , \mathbb{R} \rg $ 
invariants on a 9-point lattice either.
 Moreover instabilities close to zero lines of solutions could not
be  avoided.

To discretize the elliptic Liouville equation we proceed as in the  continuous case.  We first complexify the hyperbolic case, then we restrict to the real elliptic one. We shall need a 9-point lattice, as depicted on Fig.1 as we want to be able to discretize independently all second order partial derivatives using only invariants of the group $O(3,1)$.

We start with four points only, namely the rectangle I of Fig.1 defined by the points $(m,n), (m+1,n), (m,n+1), (m+1,n+1)$.
\begin{figure}[ht]\begin{center}\includegraphics[width=0.7\textwidth]{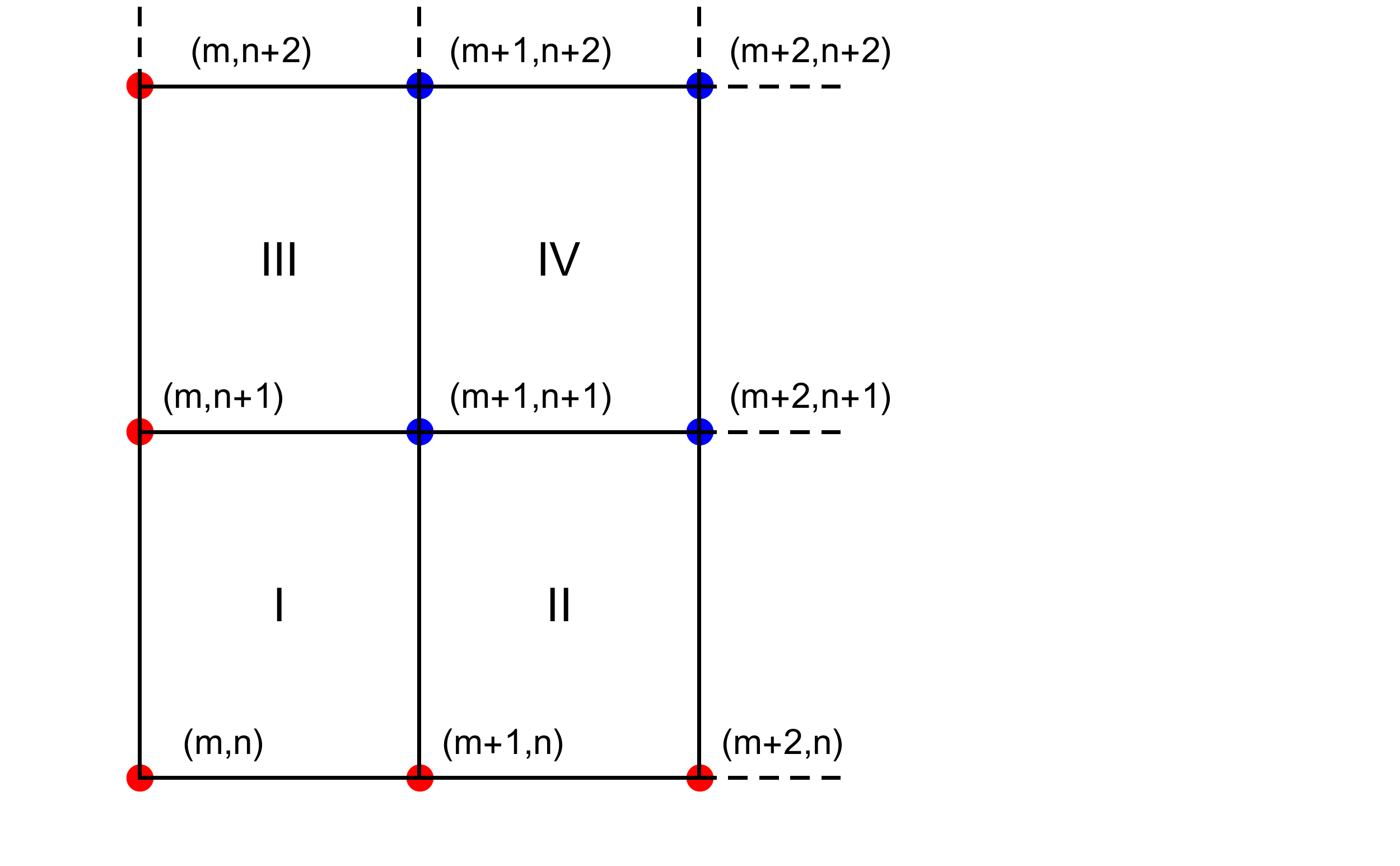}
\caption{A stencil for the 9-point scheme.} 
\end{center}\end{figure}

Six independent $ SL_x \lf 2 , \mathbb{R} \rg \otimes SL_y \lf 2 , \mathbb{R} \rg $ invariants exist on this 4-point stencil I \cite{4}, namely
\bea \label{3.1}
\xi^I&=&\frac{(x_{m,n+1}-x_{m,n})(x_{m+1,n+1}-x_{m+1,n})}{(x_{m,n}-x_{m+1,n})(x_{m,n+1}-x_{m+1,n+1})}, \\ \nonumber 
\eta^I&=&\frac{(y_{m+1,n}-y_{m,n})(y_{m+1,n+1}-y_{m,n+1})}{(y_{m,n}-y_{m,n+1})(y_{m+1,n}-y_{m+1,n+1})}, \\ \label{3.2}
H_1^I&=&u_{m,n}u_{m+1,n+1} (x_{m,n+1}-x_{m,n})^2(y_{m,n+1}-y_{m,n})^2,\\ \nonumber
H_2^I&=&u_{m+1,n}u_{m+1,n+1} (x_{m+1,n+1}-x_{m+1,n})^2(y_{m+1,n+1}-y_{m+1,n})^2,\\ \nonumber
H_3^I&=&\frac{u_{m+1,n}}{u_{m,n}}\frac{(x_{m+1,n}-x_{m,n+1})^2(y_{m+1,n}-y_{m,n+1})^2}{(x_{m,n+1}-x_{m,n})^2(y_{m,n+1}-y_{m,n})^2},\\ \nonumber
 H_4^I&=&\frac{u_{m+1,n+1}}{u_{m,n}}\frac{(x_{m+1,n+1}-x_{m+1,n})^2(y_{m+1,n+1}-y_{m+1,n})^2}{(x_{m+1,n}-x_{m,n})^2(y_{m+1,n}-y_{m,n})^2}.
 \eea
 Here  the superscript  $I$ indicates that we are in the rectangle  $I$.
 Our procedure will now be to replace $(x_{m,n},y_{m,n})$ by two complex variables $(\alpha_{m,n},\beta_{m,n})$ in all formulas and then to restrict to $\alpha_{m,n}=x_{m,n}+iy_{m,n}, \; \beta_{m,n}=\bar \alpha_{m,n}=x_{m,n}-iy_{m,n}$ with $(x_{m,n},y_{m,n})\in  \mathbb{R}^2$.
 
 We start with the space invariants $\xi$ and $\eta$:
\bea \label{3.3}
\hat \xi^I&=&\frac{[x_{m,n+1}-x_{m,n}+i\,(y_{m,n+1}-y_{m,n})]}{[x_{m,n}-x_{m+1,n}+i\,(y_{m,n}-y_{m+1,n})]}\cdot \\ \nonumber &\cdot&\frac{[x_{m+1,n+1}-x_{m+1,n}+i\,(y_{m+1,n+1}-y_{m+1,n})]}{[x_{m,n+1}-x_{m+1,n+1}+i\,(y_{m,n+1}-y_{m+1,n+1})]},\quad  \hat \eta^I=\frac 1 {\hat \xi^I}.
\eea
We see that $\hat \xi^I$ and $\hat \eta^I$ are not independent. When restricting to the real elliptic case we shall  use $\mathrm{Re}\,  \hat \xi^I$ and $\mathrm{Im} \, \hat \xi^I$ as the new invariants. Let us transform $\hat \xi^I$ to polar coordinates, defining
\bea \label{3.4}
x_{m,n}=r_{m,n} \cos \theta_{m,n}, \qquad y_{m,n}=r_{m,n} \sin \theta_{m,n}, 
\eea
We obtain 
\bea \label{3.5}
\hat \xi^I&\equiv& \frac{Re N }{\Delta} +i\, \frac{Im N }{\Delta} ,\\ \nonumber  \Delta&=&[r_{m,n}^2+r_{m+1,n}^2-2r_{m,n}r_{m+1,n} \cos(\theta_{m,n}-\theta_{m+1,n})] 
\\ \nonumber &\cdot&[r_{m,n+1}^2+r_{m+1,n+1}^2-2r_{m,n+1}r_{m+1,n+1} \cos(\theta_{m,n+1}-\theta_{m+1,n+1})], \\ \nonumber
Re N&=&(r_{m,n}^2r_{m+1,n+1}^2+r_{m,n+1}^2r_{m+1,n}^2)\\ \nonumber &+&2(r_{m,n}r_{m+1,n+1}r_{m,n+1}r_{m+1,n}\cos(\theta_{m,n}+\theta_{m+1,n+1}-\theta_{m+1,n}-\theta_{m,n+1})\\ \nonumber
&+&(r_{m,n+1}^2+r_{m,n+1}^2)r_{m,n}r_{m+1,n+1}\cos(\theta_{m+1,n+1}-\theta_{m,n})\\ \nonumber &-&(r_{m,n+1}^2+r_{m,n}^2)r_{m+1,n}r_{m+1,n+1}\cos(\theta_{m+1,n+1}-\theta_{m+1,n})\\ \nonumber
&-&(r_{m+1,n}^2+r_{m+1,n+1}^2)r_{m,n}r_{m,n+1}\cos(\theta_{m,n+1}-\theta_{m,n})
\\ \nonumber &+&(r_{m,n}^2+r_{m+1,n+1}^2)r_{m,n+1}r_{m+1,n}\cos(\theta_{m+1,n}-\theta_{m,n+1})\\ \nonumber
&-&(r_{m,n+1}^2+r_{m+1,n+1}^2)r_{m,n}r_{m+1,n}\cos(\theta_{m+1,n}-\theta_{m,n})
\\ \nonumber &-&(r_{m+1,n}^2+r_{m,n}^2)r_{m,n+1}r_{m+1,n+1}\cos(\theta_{m+1,n+1}-\theta_{m,n+1}),\\ \nonumber
Im N&=&(r_{m,n+1}^2-r_{m+1,n}^2)r_{m,n}r_{m+1,n+1} \sin(\theta_{m+1,n+1}-\theta_{m,n})\\ \nonumber &+&(r_{m+1,n}^2-r_{m+1,n+1}^2)r_{m,n}r_{m,n+1} \sin(\theta_{m,n+1}-\theta_{m,n})\\ \nonumber
&+&(r_{m+1,n+1}^2-r_{m,n+1}^2)r_{m,n}r_{m+1,n} \sin(\theta_{m+1,n}-\theta_{m,n})\\ \nonumber &+&(r_{m,n}^2-r_{m,n+1}^2)r_{m+1,n}r_{m+1,n+1} \sin(\theta_{m+1,n+1}-\theta_{m+1,n})\\ \nonumber
&+&(r_{m+1,n}^2-r_{m,n}^2)r_{m,n+1}r_{m+1,n+1} \sin(\theta_{m+1,n+1}-\theta_{m,n+1})\\ \nonumber &+&(r_{m,n}^2-r_{m+1,n+1}^2)r_{m,n+1}r_{m+1,n} \sin(\theta_{m+1,n}-\theta_{m,n+1}).
\eea
The  new real  invariants  defined by
\bea \label{3.6}
\sigma=\frac{-Re N}{\Delta}, \qquad \tau =\frac{Im N}{\Delta},
\eea
 make it possible to introduce the invariant lattice equations of type (\ref{a3}) by putting 
\bea \label{3.7}
\sigma=A,\qquad \tau=B,
\eea
where $A$ and $B$ are real constants (not depending on $m$ or $n$).  

For the hyperbolic case we simplified the problem,  by requiring that $x$ and $y$ depend on one discrete variable each: $x_{m,n}\equiv x_m, \, y_{m,n}\equiv y_n$. This implied $\xi^I=\eta^I=0$ and provided an orthogonal lattice in Cartesian coordinates.

Here, in the elliptic case, we make a similar requirement, however in the polar coordinates, namely
\bea \label{3.8}
r_{m,n}=r_m, \quad \theta_{m,n}=\theta_n.
\eea
For the lattice invariant equation (\ref{3.7})    this implies
\bea \label{3.11}
\sigma \equiv  4 \frac{r_m r_{m+1}}{(r_m-r_{m+1})^2} \sin^2\frac{\theta_{n+1}-\theta_n} 2 = A,\quad  \tau \equiv 0.
\eea
 A solution of system (\ref{3.11}) provides a class of    specific polar coordinate lattices, namely
\bea \label{3.12}
r_m=r(s+1)^m, \; \theta_n=\theta_0 + \frac{2 \pi} N n, \;n=0,1,\dots,N-1, \;s>-1, \,  N\in \mathbb{N}^*.
\eea
Thus the angle $2 \pi$ is divided into $N$ equal sectors and the radius spacing increases $(s>0)$ or decreases $(-1< s <0)$ exponentially for $m \rightarrow \infty$.  As mentioned in the Introduction the equations (like \eqref{3.7}) are invariant. The explicit lattice \eqref{3.12}, as a solution, is not invariant under $ O(3,1)$ (because we imposed the non invariant condition \eqref{3.8}). 
On the lattice determined by (\ref{3.12}),  (\ref{3.11}) reduces to
\bea \label{3.13}
A=4 \frac {s+1}{s^2} \sin^2 \frac{\pi}{N}>0.
\eea
We are thus left with four free parameters, $r$, $\theta_0$,  $-1 < s$ and $N \in  \mathbb{N}^*$.

The {\it  elliptic} invariants in (\ref{3.2}) reduce to:
\eqs{\begin{array}{l}
H^I_1 =  16 r^4 (s+1)^{4 m} \sin
   ^4(\frac{\pi
   }{N}) u_{m,n}
   u_{m,n+1} \\
H^I_2 = 16 r^4 (s+1)^{4 m+4} \sin
   ^4(\frac{\pi
   }{N}) u_{m+1,n}
   u_{m+1,n+1} \\
H^I_3 =\Big [  \frac{ s^2 +4(s+1)\sin^2(\frac{\pi
   }{N}) }{4 \sin^2 (\frac{\pi
   }{N})
  }\Big]^2 \frac{ u_{m+1,n}}{ u_{m,n}} \\
H^I_4 =  \frac{16 (s+1)^4 
  }{s^4}\sin
   ^4(\frac{\pi
   }{N})\frac{ u_{m+1,n+1}}{ u_{m,n}} 
\end{array}\label{3.14}}

The continuous limit corresponds to
\bea \label{3.16}
N \rightarrow \infty, \qquad s \rightarrow 0,\qquad  \frac n N \rightarrow \frac{\theta}{2 \pi} \qquad sN=2 \pi .
\eea
The condition $s N=\kappa$, with $\kappa$ finite, follows from the fact that $A$ of (\ref{3.13}) is finite. The choice $\kappa=2 \pi$ is required, as we shall see below in (\ref{3.19}), in order to obtain the correct { continuous} limit for the discrete equation.

In order to write a discrete equation,  with the correct  continuous limit, we introduce a more convenient set of invariants, with $\sigma$, $\tau$ as in (\ref{3.6}). They are:
\eq{J^I_1 =  \frac{H^I_4}{\sigma^2} = (s+1)^2\frac{ u_{m+1,n+1}}{u_{m,n}}, \qquad J_2^I = \frac{\sigma^2 H^I_2}{4 H^I_1 H^I_4} =(s+1)^2 \frac{ u_{m+1,n}}{u_{m,n+1}} .\label{3.15}}

Furthermore,  by using (\ref{3.12}) we have 
\bea \label{3.17}
u_{m+j,n+k}=u(r(s+1)^{m+j},\frac{2 \pi} N (n+k)), \quad 0 \le j \le 2, \; 0 \le k\le2. 
\eea
Let us define
\bea \label{3.18}
I_1= -J^{III}_1 + J^{IV}_1 + J^{II}_2 - J^{IV}_2 , \quad
I_2=\sum_{i=1}^2{a_i \sqrt{H^I_i }} + \sum_{i=3}^4{a_i \sqrt{H^I_1 H^I_i }},
\eea
where $J^{II}_i ,\; J^{III}_i, \; J^{IV}_i$,  $i = 1,2$, are obtained  by shifting the invariants $J_1^I$ and $J_2^I$  in the corresponding region of the 9-point lattice. The  expansion of $u_{m+j,n+k}$ in $s$ and $\frac{2 \pi} N$ about $u_{m,n} \equiv u(r,\theta)$, keeping derivatives up to order 2, leads to the following  expansion  for $I_1$ and $I_2$  in terms of $s$ and $\frac 1 N$
\bea \label{3.19}
 I_1 &=& \frac 1 {u^2} \{ s^2 [ r^2 (u u_{rr} -u_r^2) + r u u_r ] + (\frac{2 \pi} N)^2 [u u_{\theta \theta} - u_{\theta}^2]\} +\\ \nonumber &+&\mathcal O(s^3, (\frac{2 \pi} N)^3,s^2 \frac{2 \pi} N, s (\frac{2 \pi} N)^2), \\ \nonumber
 I_2 &=& u r^2 \{ (\frac{2 \pi} N)^2 [a_1 + a_2 +a_3] + s^2 a_3 + \frac{(\frac{2 \pi} N)^4}{s^2} a_4 \}+\\ \nonumber &+&\mathcal O(s^3, (\frac{2 \pi} N)^3,s^2 \frac{2 \pi} N, s (\frac{2 \pi} N)^2).
\eea
If we choose  $sN=2\pi$, as in (\ref{3.16}),  the lowest order term in the expansion  of $I_1$ will be equal to $ (\frac{2 \pi} {N u})^2$ times the left hand side of (\ref{2.9}). The lowest order term in the expansion of $I_2$ will be $ (\frac{2 \pi} {N u})^2$ times the right hand side of (\ref{2.9}) once we put $a_1+a_2+2a_3+a_4=1$ .

Hence an $ O(3,1)$ invariant scheme  on the lattice specified by (\ref{3.12}) is obtained by requiring 
  \eqs{I_1=I_2 . \label{InvCond}} Setting for simplicity $ a_3 = a_4 = 0$, i.e. $a_1+a_2=1$,  (\ref{InvCond})  reads
 \eqs{&&-4 r^2 (s+1)^{2 m} \sin^2(\frac s 2) [ a_1 \sqrt{u_{m,n}
   u_{m,n+1}}+ (1-a_1)  (s+1)^{2}
   \sqrt{u_{m+1,n}
   u_{m+1,n+1}}]\nonumber\\ &&\label{3.20}+ (s+1)^2 [-\frac{
   u_{m+1,n+2}}{u_{m,n+1}}+ \frac{
   u_{m+2,n+2}}{u_{m+1,n+1}}+ \frac{
   u_{m+2,n}}{u_{m+1,n+1}}-\frac{ u_{m+2,n+1}}{u_{m+1,n+2}} ]= 0 .  }
Eq. (\ref{3.20}) is the discrete elliptic Liouville equation, invariant under the group $O(3,1)$, {together with an expression of the  lattice $r_m$, $\theta_n$ \eqref{3.12}. It can be used to calculate $u_{m+2,n+2}$, once $u_{m+j,n+k}$ is known for all other points on the 9-point stencil (actually $u_{m,n+2}$ is not used). The continuous limit is, to the lowest order (i.e. $s^2$) 
 \eq{\frac{
   (r (r
   u_{r,r}+u_r)+u_{\theta,\theta}) u-r^2
   u_r^2-u_{\theta}^2}{u^2}=r^2 u + \mathcal O(s).\label{limits}}  
  The term $\mathcal O(s)$ is very involved, depending on $u(r, \theta)$ and its derivatives,  but the coefficients depend explicitly on   the discrete label $m$ of the lattice. This gives secular terms
and thus makes the above limit non-uniform.

It is worthwhile to mention  that by shifting the invariants (\ref{3.1}, \ref{3.2})  to the rectangles $II$, $III$  and $IV$
on Fig.1 we have obtained a complete set of 24 functionally independent $O(3,1)$ 
invariants. We are however not able to construct a difference scheme out
of them that is invariant under the entire infinite-dimensional
conformal group of the plane $E_2$.
\section{Conclusions}
The main result presented in this article is the invariant discrete
Liouville equation \eqref{3.20} on the lattice \eqref{3.12}. Both the equation and the
lattice are solutions of the difference scheme
$I_1 = I_2 , \quad \sigma = A, \;  \tau = 0$
with $I_1$, $I_2$ , $\sigma$ and  $\tau $ defined in \eqref{3.18} and \eqref{3.6}, respectively.
This scheme is invariant under $O(3,1)$, the maximal finite-dimensional
subgroup of the infinite-dimensional conformal group of the Euclidean
plane.    
 It turns out that we need a  9-point lattice to be able to reproduce the two second order derivatives. This implies that, as pointed out in \cite{5} for the hyperbolic Liouville equation,   boundary conditions are more difficult to impose. Moreover the continuous limit shows the presence of secular terms which might lead to instabilities in the numerical results.

{{ Numerical calculations of solutions of the {  hyperbolic} Liouville equation  were presented in ref. \cite{4} and \cite{5} and  the conclusions were stated above. For the {elliptic} Liouville equation the situation is different: the only discretization that we can compare to is the standard one or nonstandard  difference schemes \cite{b,m}. To our knowledge no discretization preserving the infinite-dimensional symmetry group of point symmetries as generalized symmetries or preserving linearizability exists. Moreover, as we must have an eight point discretization, instabilities close to zero lines of solutions cannot
be  avoided. 


 Lie group theory has its most powerful applications to partial differential equations, specially when the groups are  infinite-dimensional. A study of all possible extensions of this { approach}  to  multivariable discrete equations would be most appropriate{, possibly using the formalism developed in}  \cite{a9,a10,a11,a12}.

\paragraph{Acknowledgments.}   DL and LM have been partly supported by the Italian Ministry of Education
and Research, 2010 PRIN {\it Continuous and discrete nonlinear integrable evolutions: from
water waves to symplectic maps} and by INFN IS-CSN4 {\it Mathematical Methods of Nonlinear
Physics}. The research of PW is partially supported by an NSERC discovery grant.


\end{document}